\documentclass[aps,prl,preprint,superscriptaddress]{revtex4}
\pdfoutput=1
\usepackage{graphicx}
\usepackage{dcolumn}
\usepackage{bm}
\usepackage{color}
\usepackage[thinspace,thinqspace,amssymb,pstricks]{SIunits}
\usepackage{pslatex}
\usepackage{color}
\begin{document}

\title{Fermi surfaces of single layer dielectrics on transition metals}

\author{T. Greber, M. Corso$^*$, and J. Osterwalder}
\affiliation{Physik-Institut, Universit\"{a}t Z\"urich, Winterthurerstrasse 190, CH-8057 Z\"{u}rich, Switzerland\\
$^*$ present address: DIPC, Manuel de Lardizabal 4, E-20018 San Sebastian, Spain}
\date{\today}

\begin{abstract}

Single sheets of hexagonal boron nitride on transition metals provide a model system for single layer dielectrics. 
The progress in the understanding of h-BN layers on transition metals of the last 10 years are shortly reviewed. 
Particular emphasis lies on the boron nitride nanomesh on Rh(111), which is a corrugated single sheet of h-BN, where the corrugation imposes strong lateral electric fields.
Fermi surface maps of h-BN/Rh(111) and Rh(111) are compared. A h-BN layer on Rh(111) introduces no new bands at the Fermi energy, which is expected for an insulator. The lateral electric fields of h-BN nanomesh violate the conservation law for parallel momentum in photoemission and smear out the momentum distribution curves on the Fermi surface.\\
\\
\end{abstract}

\small{\it{Accepted for publication in the Surface Science special issue in honor of Gerhard Ertl $'$s Nobel Prize}}

\maketitle

\begin{figure}[b]
	\begin{center}
	\includegraphics[width=0.5\textwidth]{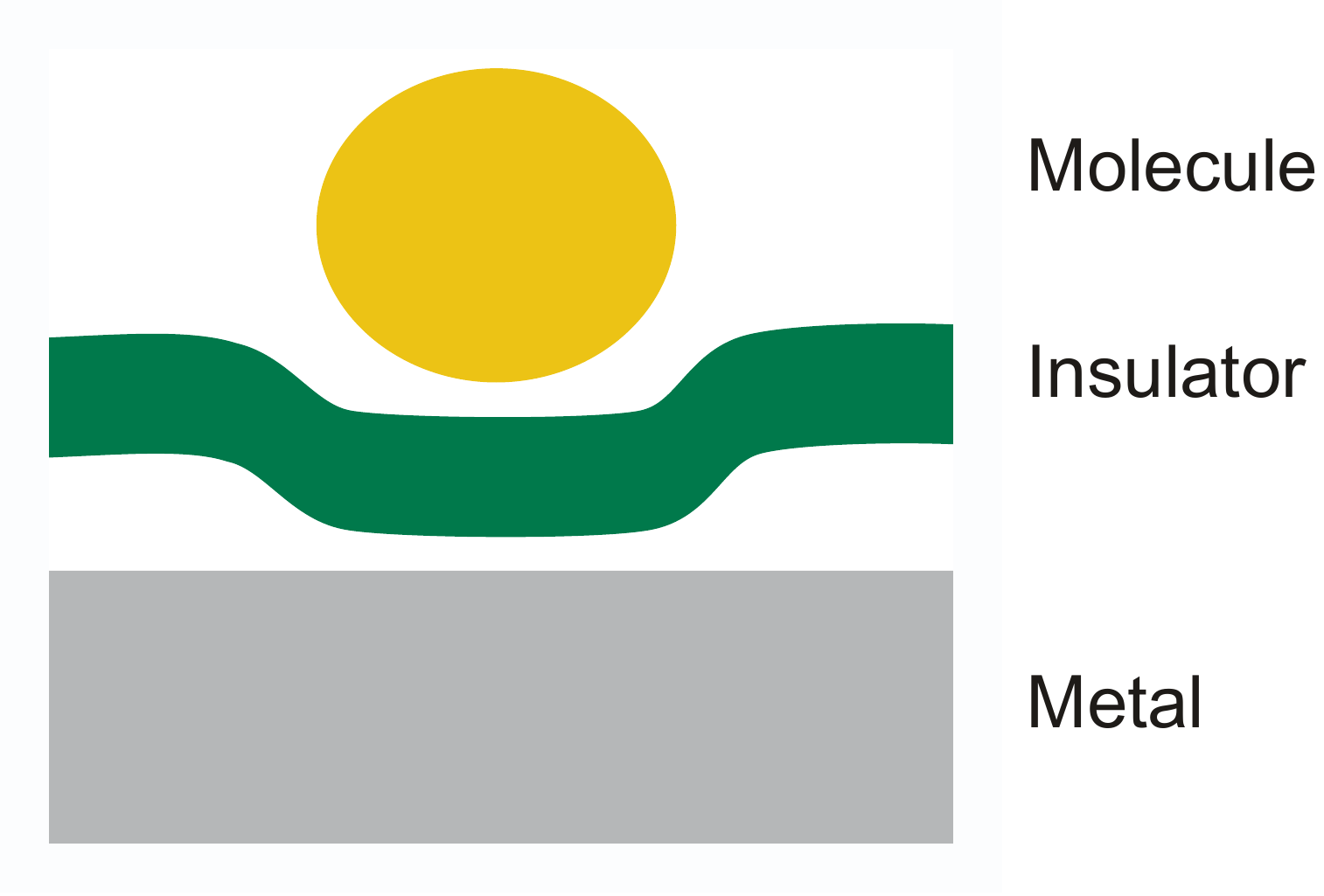}
		\caption{(Color online) Schematic side view of a single layer dielelectric (insulator) on top of a metal. The dielectric strongly changes the properties of adsorbed molecules as compared to the case without dielectric. If the dielectric corrugates, the structures may be used as templates on which single molecules may be laterally ordered.}
		\label{F1}
	\end{center}
\end{figure}

The understanding of the principles of mobility of charge is one of the most fundamental issues of chemistry and physics. 
These principles rule processes like molecular bond formation and bond breaking, or those of electrical conductivity. 
For the investigation of charge mobility surface science plays a key role because it allows very direct insight at the atomic level \cite{ert94}.
Its success foots on the investigation of simple, but non trivial model systems, though in technically demanding conditions. 
The interpretation of its experimental results enable the bottom up understanding of charge transfer processes. 
Here a model system for such studies, i.e. a single layer of hexagonal boron nitride on transition metals, is discussed.
It is a substrate which is a hybrid between a reactive transition metal surface and an inert wide band gap insulator, or dielectric.
One monolayer of a dielectric is not sufficient to inhibit charge contact between the metal and any adsorbate, due to the extended volume of the electron wave functions and tunneling contact is expected. 
Furthermore, the dielectric layers lower the adsorption energy of molecules \cite{ram06}.
These two aspects make single layer dielectrics surface systems with new functionalities, where e.g. molecular adsorbates change their vacuum properties to a much smaller extent than they do on a transition metal (see Figure \ref{F1}).

There are many  single layer systems like e.g.  graphite/graphene \cite{she74,ndi06}, hexagonal boron nitride \cite{paf90,auw99},  boron carbides \cite{yan04}, molybdenum disulfide \cite{hel00}, sodium chlorides \cite{ben99,piv05} or aluminium oxide \cite{nil08}, to name a few. 
In order to decide whether single layers were "dielectric" or "metallic" the electronic structure at the Fermi level has to be studied, where a metallic layer introduces new bands at the Fermi energy, while a dielectric layer does not.

When the first h-BN single layers were synthesized by a chemical vapor deposition (CVD) process that comprised the reaction of benzene like borazine $(HBNH)_3$ with  a clean Ru(0001) and Pt(111) surface, it was argued from the analysis of the low energy electron diffraction (LEED) pattern  and Auger electron emission spectroscopy that single layer super lattices were formed \cite{paf90}. 
The CVD process comprises the hydrogen abstraction from the borazine molecules, the assembly of hexagonal boron nitride and the desorption of $H_2$ gas.
The h-BN growth rate drops after the formation of the first layer by several orders of magnitude.
This has the practical benefit that it is easy to prepare single layers.
The drop in growth rate is presumably due to the much lower sticking coefficient of borazine after the completion of the first h-BN layer, and the much lower catalytic activity of h-BN compared to clean transition metals. 
However, not much is known on the details of the growth process of h-BN.
From the study of h-BN island morphologies on Ni(111) it was suggested that the borazine BN six-ring is opened during the self assembly process \cite{auw03,auw04}.  

The lattice constants of the h-BN/Pt(111) and the h-BN/Ru(0001) superlattices were one order of magnitude larger than those of the 1x1 unit cells and were rationalized with the lattice mismatch between h-BN and the transition metal. 
The size of the super cell can be determined from $N\cdot a_{TM}=(N+1)\cdot a_{hBN}$, where $N$ is the number of substrate 1x1 transition metal (TM) lattice constants  $a_{TM}$, and $a_{hBN}$ the adopted lattice constant of h-BN. 
Generally, $N$ is not an integer number. 
The difference to the next integer may cause incommensurability or lock in to an integer number, commensurate, $N \times N / (N+1) \times (N+1)$ coincidence lattice. 
For the case of h-BN on Pt(111) and Ru(0001) it is believed that the energy cost for the residual straining into the commensurate super lattice is smaller than the gain of lock in energy. 
The residual straining in order to get coincidence is an order of magnitude lower and accordingly costs two orders of magnitude less energy.
Correspondingly, 10x10 Pt/11x11 h-BN, and 12x12 Ru/13x13 h-BN superstructures were inferred. 
Since the h-BN layers are grown at high temperatures, it also has to be considered that the differences in lateral thermal expansion of the substrate and the adsorbate cause for Ru(0001) at 1000ÊK a  $N$ of 11.5 and at room temperature a $N$ of 12.4. 

Later on, single layer h-BN was grown on Ni(111) \cite {nag952}. Here $a_{Ni}$ is 0.4\% smaller than $a_{hBN}$ and 1x1 structures have been reported \cite{gam97}; and from photoemission it was concluded that h-BN physisorbs \cite{nag951}.
The application of scanning tunneling microscopy to h-BN/Ni(111) showed that indeed perfect single domain single layers are formed \cite{auw99}. 
This substrate was further explored in view of its usefulness for building metal insulator metal or molecule interfaces. 
It was e.g. shown that the sticking coefficient of cobalt atoms is temperature dependent and surprisingly small \cite{auw02}.
Further experiments that investigated the charge state of $C_{60}$ on h-BN/Ni(111) found an intriguing 700\% increase of the charge state of $C_{60}$, in a phase transition between 200 and 300ÊK that was  assigned to the onset of  $C_{60}$ rocking motion \cite{mun05}. 
This observation of charge transfer can qualitatively be explained by the difference in tunneling matrix element in changing the orientation of  $C_{60}$ on h-BN/Ni(111).
However, the magnitude of the effect can not be understood in an adiabatic picture \cite{mun05,che05}.
The proposed mechanism for the effect is traced back to a kind of electron self trapping on  $C_{60}$.
This effect is only sizeable due to the relatively low tunneling rate of conduction electrons  across a single layer h-BN, where timescales of molecular and electronic motion start to match. 
If an electron tunneled from the substrate to the molecule, the molecule accelerates towards the surface. 
Since the tunneling rate is low, the residence time of the electron is long and the displacement of the molecule becomes sizeable.
This causes the electron on the affinity level, or the so called lowest unoccupied molecular orbital (LUMO) of the $C_{60}$, to dive below  the Fermi level of the metal substrate. As a consequence the back-tunneling is further suppressed due to the unavailability of empty states below the Fermi level \cite{mun05}. 
The basic concept for this explanation of the large charge transfer effect in this phase transition of the  $C_{60}$/h-BN/Ni(111) system has been borrowed from those in non-adiabatic gas-surface reactions \cite{bot90,gre973}.

\begin{figure}
	\begin{center}
	\includegraphics[width=0.55\textwidth]{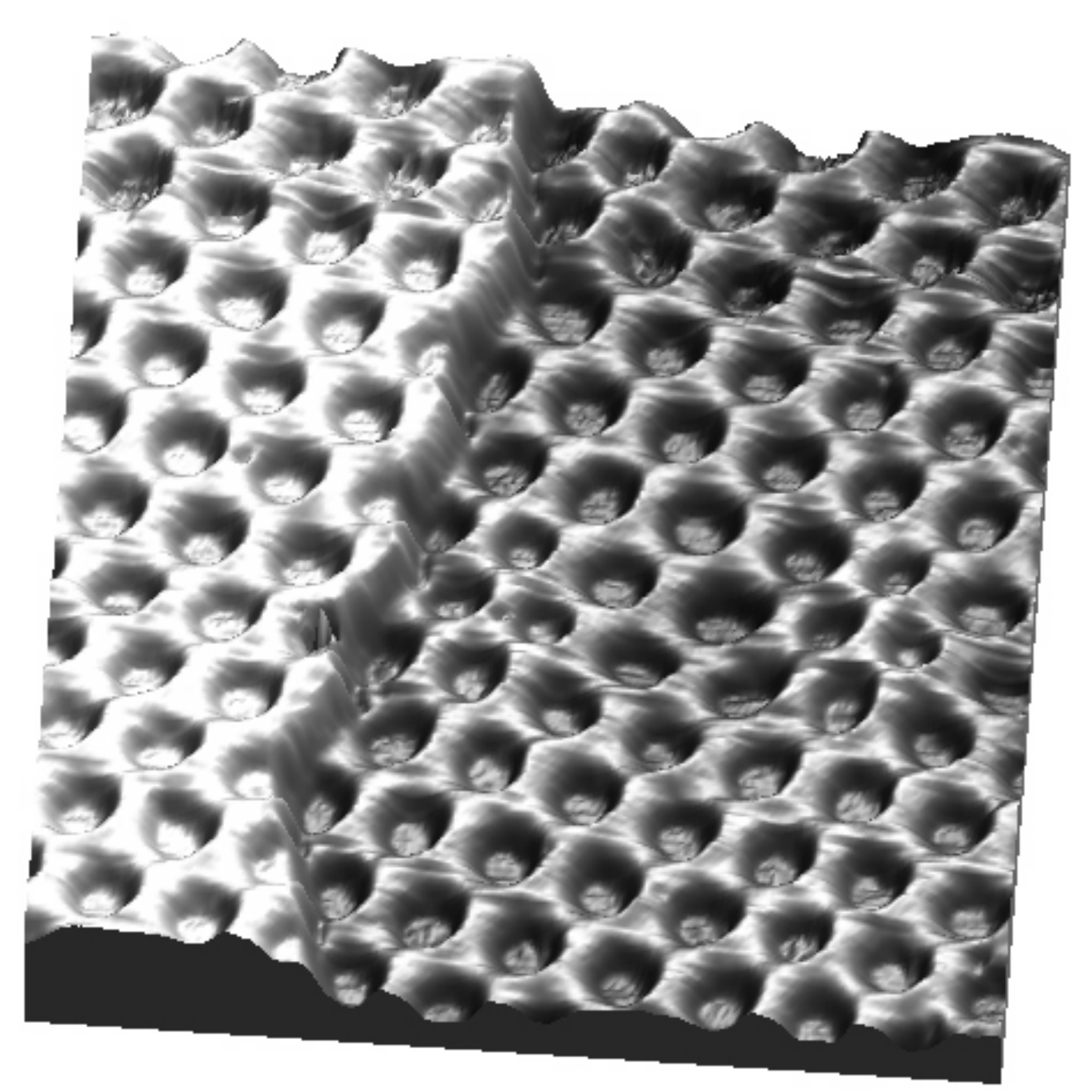}
	
		\caption{ Constant current STM image of boron nitride nanomesh (h-BN/Rh(111)) (30x30 nm$^2$, I$_t$=2.5 nA, V$_s$=-1 V). This honeycomb nanostructure with 3.2+0.1 nm periodicity consists of two distinct areas: the wires which are 1.2+0.2nm broad and the holes with a diameter of  2.0+0.2nm. The corrugation is 0.7+0.2 \AA.}
		\label{FSTM}
	\end{center}
\end{figure}

On the search for a better understanding of the h-BN layers, also Rh(111) substrates were investigated, where a distinct structure was discovered and subsequently named nanomesh \cite{cor04}. 
A multi method approach, including low energy electron diffraction (LEED), scanning tunneling microscopy (STM) and photoemission, helped to establish this nano structure. 
LEED indicated a 12x12 Rh/13x13 h-BN coincidence lattice which corresponds to a lattice constant of about 3 nm.
STM suggested that the superlattice could be explained by a moir\'e type pattern, if  a moir\'e pattern is understood as the superposition of two two-dimensional lattices with different lattice constants or orientations, which causes a kind of "beating" in the lateral dependence of the tunneling current \cite{ron93}. 
Figure \ref{FSTM} shows a STM picture of h-BN nanomesh across a monoatomic step of the Rh(111) substrate. A honeycomb array of "holes" with 2 nm diameter, separated by 3.2 nm displays.
The decoration of the h-BN/Rh(111) superstructure with $C_{60}$ illustrated the potential of the nanomesh for the formation of supramolecular structures.
Valence band photoemission indicated a h-BN $\sigma$ band splitting of about 1 eV \cite{cor04}. This was new and stood in contrast to h-BN/Ni(111) \cite{gra03,nag951}, or to h-BN/Pd(111) that forms rotational moir\'e  patterns \cite{mor06}, but has no $\sigma$ band splitting \cite{nag951,mor06}.

\begin{figure}
	\begin{center}
	\includegraphics[width=0.9\textwidth]{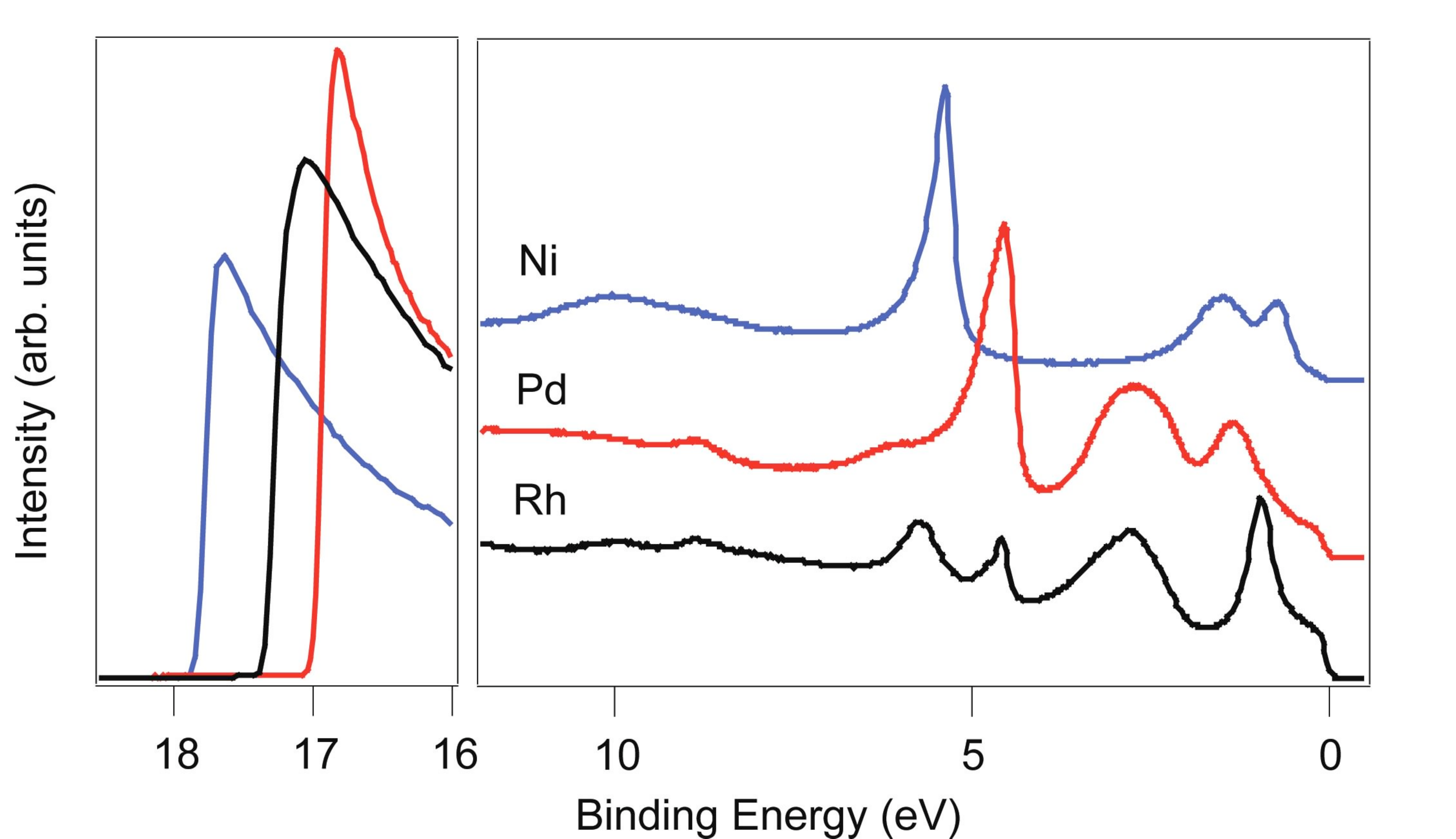}

		\caption{(Color online) He $I_\alpha$ normal emission photoemission spectra of h-BN/Ni(111), h-BN/Pd(111) and h-BN/Rh(111). While h-BN/Ni(111) \cite{gre02} and h-BN/Pd(111) show no sizeable $\sigma$ band splitting, the splitting into $\sigma_\alpha$ and $\sigma_\beta$ for h-BN/Rh(111) is about 1ÊeV \cite{cor04}.}
		\label{F2}
	\end{center}
\end{figure}

Figure \ref{F2} shows valence band photoemission data for h-BN/Ni(111), h-BN/Pd(111) and h-BN/Rh(111). 
Hexagonal boron nitride has 3 $\sigma$ bands and one $\pi$ band that are occupied.
Along $\bar{\Gamma}$, i.e. perpendicular to the $sp^2$ plane, the two low binding energy $\sigma$ bands are degenerate, while the third is not accessible for He I$_\alpha$ radiation.
The $\sigma$ bands of h-BN/Ni(111) and of h-BN/Pd(111) degenerate into a single peak. Those of h-BN/Rh(111) are weak and split in a $\sigma_\alpha$ and a $\sigma_\beta$ contribution \cite{gor07}. 
In measuring the spectra away from the $\bar{\Gamma}$ point it is seen that both, the $\sigma_\alpha$ and the $\sigma_\beta$ bands split in two components each, as it is known from angular resolved measurements of h-BN single layers with $\sigma_\alpha$ bands only \cite{nag951}.

\begin{table}
\caption{Experimental values (in eV) for the photoemission binding energies referred to the Fermi level $E_B^F$ of the $\sigma$ bands for h-BN single layers on three different substrates along $\bar{\Gamma}$, and the corresponding work functions $\Phi$. The binding energies with respect to the vacuum level $E_B^V$ are determined by $E_B^F+\Phi$.\\}
 \begin{tabular}{c|c|c|c|c|c}
 \hline
 Substrate&$E_B^FÊ\sigma_\alpha$&$E_B^FÊ\sigma_\beta$&$\Phi$&$E_B^VÊ\sigma_\alpha$&Reference\\
 \hline
 Ni(111)&5.3& &3.5&8.8&\cite{gre02}\\
 Rh(111)&4.57&5.70&4.15&8.7&\cite{gor07}\\
 Pd(111)&4.61&&4.26&8.9&this work\\
  \hline
 \end{tabular}

\label{T1}
\end{table}

The  $\sigma$ band splitting indicated two electronically different regions within the nanomesh unit cell.
They were related to strongly bound and loosely bound h-BN. 
Later on, h-BN/ Ru(0001) \cite{gor07}  was found to be very similar to h-BN/ Rh(111) and it was shown by theoretical efforts \cite{las07} and atomically resolved low temperature STM \cite{ber07} that the h-BN/Rh(111) nanomesh is a corrugated single sheet of h-BN on Rh(111). 
The peculiar structure arises from the strong site dependence of the interaction with the substrate and causes the corrugation of the h-BN sheet, with a height difference between strongly bound regions and loosely bound regions of about 0.5Ê\AA. 
The structure releases energy in forming dislocations in the 12x12/13x13 unit cell. 
The anisotropic elasticity of h-BN favors outward relaxation that can be seen as a dislocation where a strain relief with "holes" or "pores" forms low regions and "wires" as the high regions of the nanomesh. 
The low or tightly bound regions are circular patches with about 2 nm diameter. 
In these regions the nitrogen atoms bind on top of the substrate atoms \cite{las07}, which is in line with h-BN/Ni(111) \cite{gam97,mun01,gra03}.
It was realized that the nanomesh may be used as a template for the periodic arrangement of single molecules, where it emerged as a promising construction lot for building nanostructures. 
For example, if naphthalocyanine, a molecule with a diameter of about 2 nm, is evaporated on the nanomesh, isolated molecular entities separated by 3.2 nm are found immobilized at room temperature \cite{ber07}. 
Another interesting feature of the nanomesh is its stability in air and liquids \cite{ber07}.  The stability of the h-BN nanomesh superstructure in an electrolyte was best documented by in situ STM images  \cite{wid07}.
 
\begin{figure}
	\begin{center}
	\includegraphics[width=0.5\textwidth]{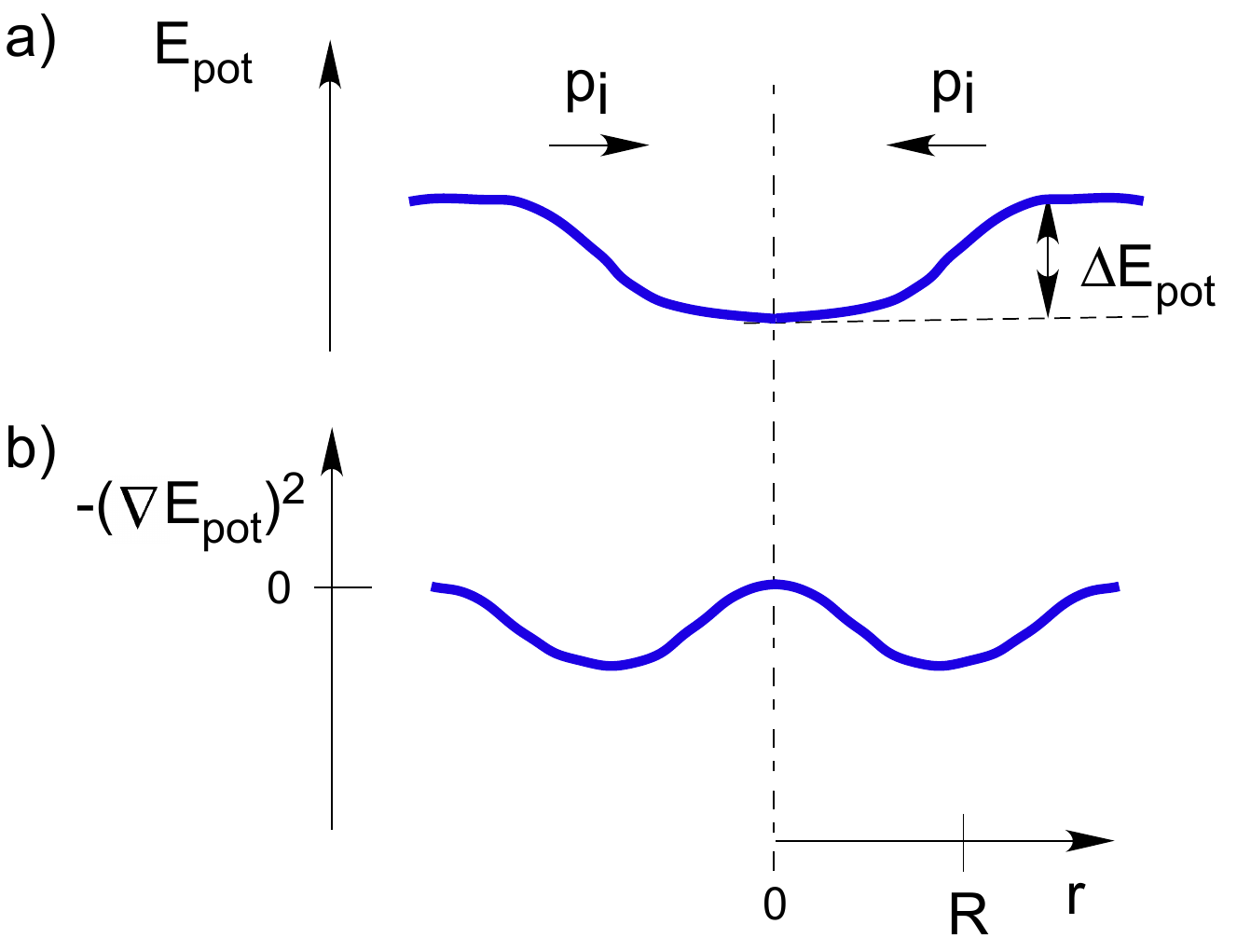}
		\caption{(Color online) Schematic drawing of a dipole ring where in plane dipoles $\textbf{p}_i$ are sitting on a ring with radius $R$. a) the potential energy $E_{pot}$ and b) the polarization energy, which is proportional to $(\nabla E_{pot})^2$ in dipole rings.}
		\label{F3}
	\end{center}
\end{figure}

 It took some more time to improve the understanding of the distinct electronic structure and the concomitant functionality as a trap for molecules, which match the size of the holes.
The key for the understanding lies in the $\sigma$ band splitting.
This 1 eV splitting is also reflected in N1s core level x-ray photoelectron spectra \cite{pre072}, where the peak assignment is in line with the $\sigma$ band assignment \cite{gor07,ber07}.
Without influence of the substrate the energy of the $\sigma$ band, which reflects the in plane $sp^2$ bonds, are referred to the vacuum level. 
Vacuum level alignment arises for physisorbed systems, as e.g. for noble gases \cite{wan84,jan94}, or as it was proposed for h-BN films on transition metals \cite{nag951}.
The $\sigma$ band splitting causes the conceptual problem of aligning the  vacuum level and the Fermi level with two different work functions.
The {\it{local}} work function, or the electrostatic potential near the surface may, however, be different, and it may be determined e.g. in measuring the energy levels of adsorbed Xe \cite{kup79,wan84}. 
The method of photoemission from adsorbed Xe (PAX) was successfully applied to explore the electrostatic energy landscape of the nanomesh \cite{dil08}.
This might revive the interest in this method for the investigation of electrostatic potentials in nanostructures.
In accordance with density functional theory calculations it was found that the electrostatic potential at the Xe cores in the holes of the nanomesh is 0.3 eV lower than that on the wires.
This has implications for the functionality of the nanomesh, since these sizeable potential gradients polarize molecules and may be used as electrostatic traps for molecules or negative ions.
The peculiar electrostatic landscape in the nanomesh has been rationalized with dipole rings, where in plane dipoles, sitting on the rim of the nanomesh holes produce the electrostatic potential well in the holes.
For small in plane dipoles that sit on a ring the electrostatic potential energy difference in the center of the ring becomes
\begin{equation}
\Delta E_{pot}=\frac{e}{4\pi\epsilon_0}\frac{P}{R^2}
\label{E1}
\end {equation}

where $e$ is the elementary charge, $R$ the radius of the hole and $P=\sum |\textbf{p}_i|$ the sum of the absolute values of the dipoles on the ring. 
For $R$=1 nm and $\Delta E_{pot}$=0.3 eV, $P$ gets 10ÊD, which is equivalent to 5.4 water molecules with the hydrogen atoms pointing to the center of the holes.
The strong lateral electric fields in the BN nanomesh may be exploited for trapping molecules, or negatively charged particles, where it might also act as an array of electrostatic nanolenses for slow charged particles that approach or leave the surface.
Figure \ref{F3} shows the electrostatic potential in a dipole ring and the square of the related electric fields.
For the case of the nanomesh the origin of the in plane electrostatic fields is not the Smoluchowski effect where the delocalisation of the electrons at steps forms in plane dipoles \cite{smo41}, but is due to the contact of differently bonded boron nitride with different work functions. 

Very important for the functionality of single layer systems is the question on whether it is metallic or not metallic (dielectric).
The fact that we can tunnel electrons across single layer dielectrics does not allow an answer.
For the case of h-BN/Ni  the interpretation of phonon dispersions \cite{rok97} and high energy X-ray absorption probes \cite{pre04} inferred metallicity.
The low sticking coefficient of cobalt on h-BN/Ni(111) \cite{auw02}, the peculiar behavior of C$_{60}$ on h-BN/Ni(111) \cite{mun05} and the fact that the bond energy of Xe on h-BN/Rh(111) is in between first and second layer Xe adsorption energies \cite{dil08,ker05}, are, on the other hand, indicative for a low density of states at the Fermi level at the vacuum h-BN interface.
Certainly, metallicity is given for an overlayer, if new bands cross the Fermi level. 
This can be decided in comparing Fermi surfaces from the bare and the overlayer covered substrate.
In the case of h-BN/Ni(111) all new features in the Fermi surface maps as measured with photoemission  were assigned to be bulk derived and no new bands were found \cite{gre02}.
Also theory found no strong hybridisation between h-BN and Ni states at the Fermi level \cite{gra03}.

\begin{figure}
	\begin{center}
		\includegraphics[width=0.5\textwidth]{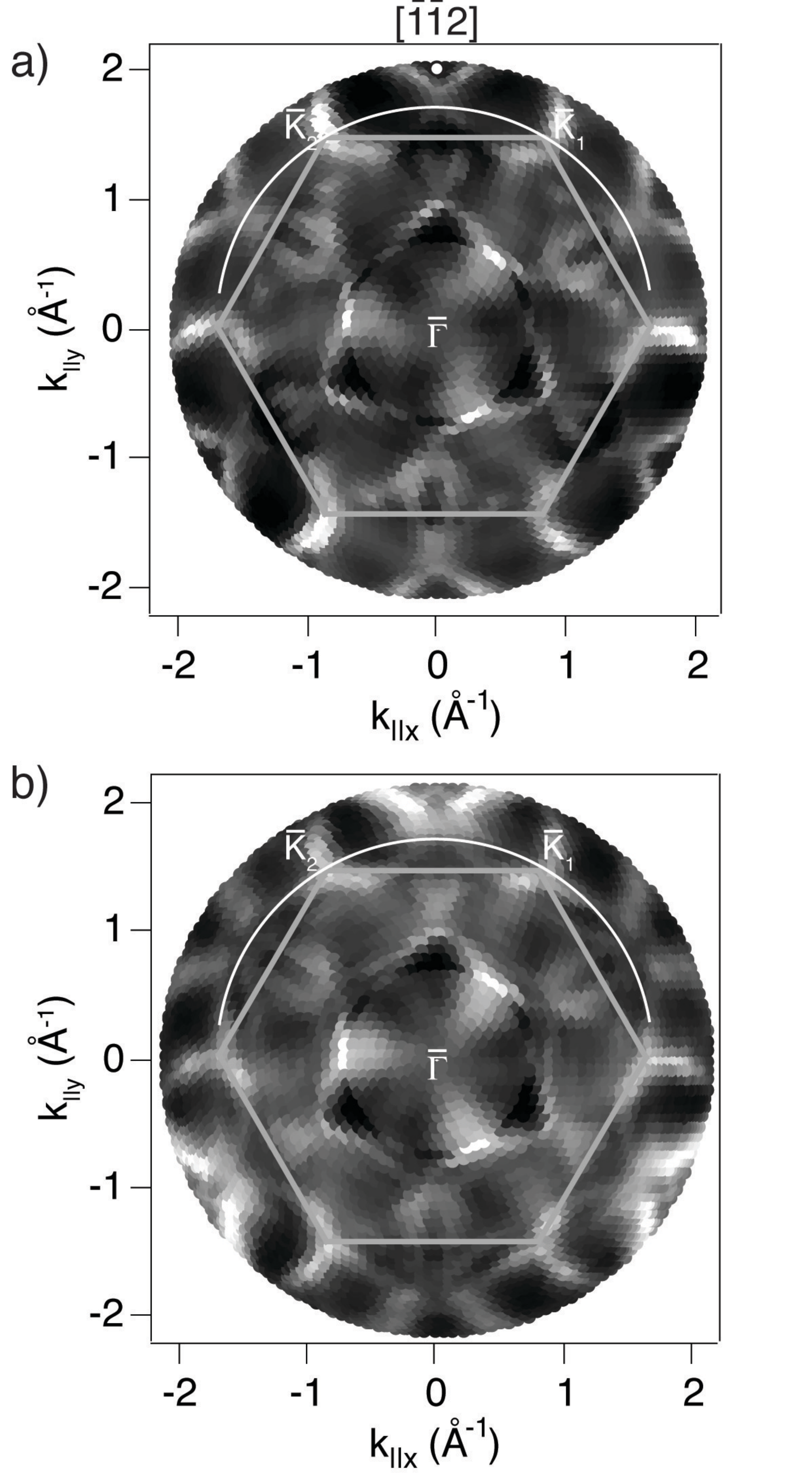}
		\caption{Fermi surface maps of a) Rh(111) and b) h-BN/Rh(111). Data were recorded with He I$_\alpha$ radiation. The intensities are averaged for each polar emission angle and displayed on a linear grey scale in $k_\parallel$ projection. The 1x1 surface Brillouin zone of h-BN and high symmetry points $\bar{\Gamma}, \bar{K}_1, \bar{K}_2$ and the azimuthal cut shown in Figure \ref{F5}c) are overlayed. Note the close resemblance between the two but the smeared appearance for the Fermi surface map of h-BN/Rh(111). }
		\label{F4}
	\end{center}
\end{figure}

\begin{figure}
	\begin{center}
		\includegraphics[width=0.7\textwidth]{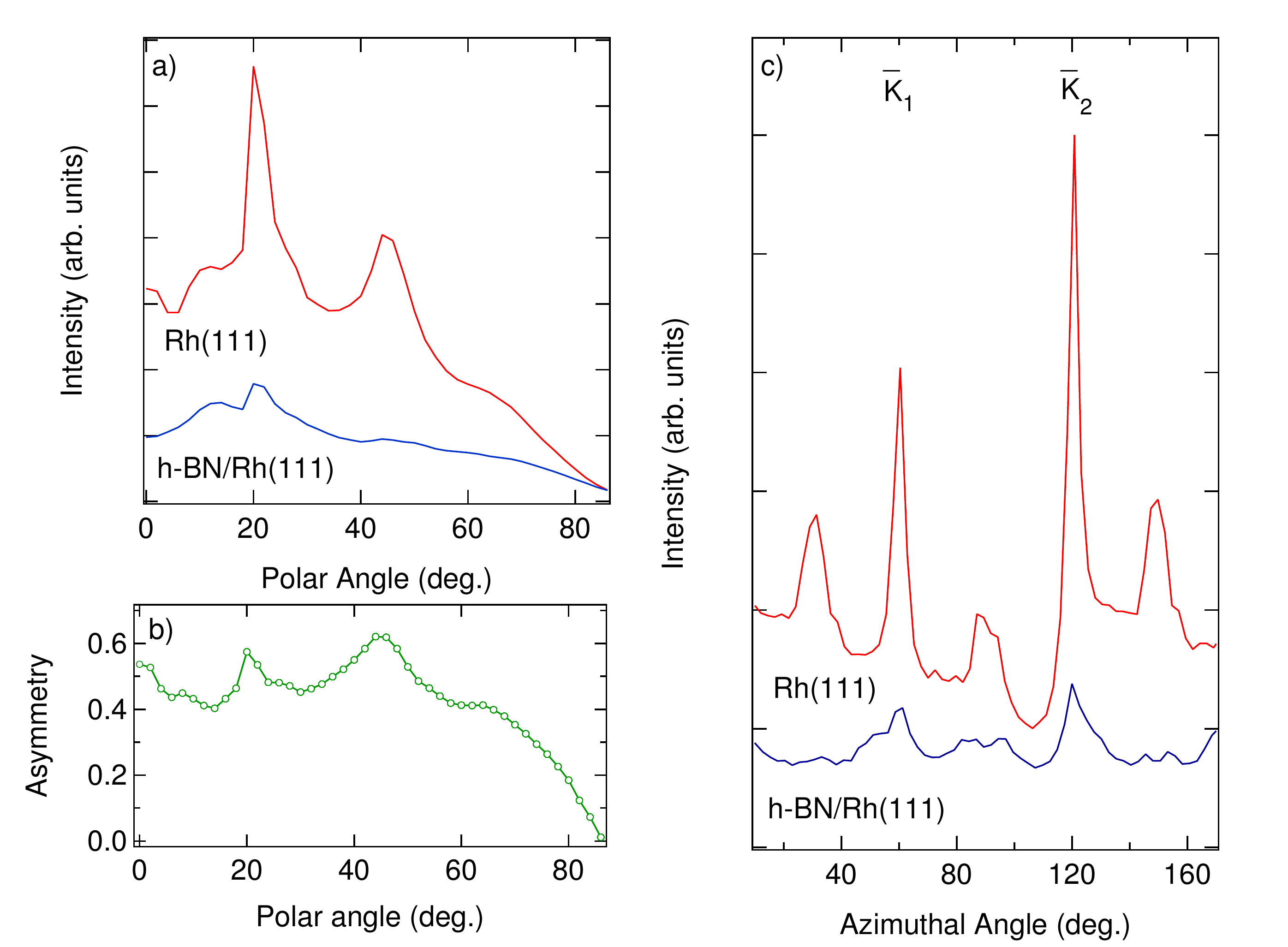}
		\caption{(Color online) Comparison of intensity variations of the two Fermi surface maps shown in Figure \ref{F4}. a) Azimuthally averaged Fermi level intensities for Rh(111) and h-BN/Rh(111) as a function of polar angle $\theta$. b) Corresponding asymmetry $A=(I_{Rh}-I_{hBN})/(I_{Rh}+I_{hBN})$. c) Momentum distribution curve on azimuthal cuts at $k_\parallel$=1.7 $\AA^{-1}$. 
		The high symmetry points $ \bar{K}_1, \bar{K}_2$ of Figure \ref{F4} are shown as well. Note the attenuation and the broadening of the peaks that correspond to Fermi level crossings of Rh bulk bands. }
		\label{F5}
	\end{center}
\end{figure}

The recording of Fermi surface maps with photoemission became a method for the investigation of the electronic structure that is most relevant for electron transport  \cite{san91,aeb94}.
In Figure \ref{F4} the Fermi surface maps of Rh(111) and h-BN/Rh(111) as measured with He I$_\alpha$ radiation are shown. 
The photoemission intensity at the Fermi level is displayed on a grey scale as a function of the electron wave vector component parallel to the surface $k_\parallel$.
The 1x1 surface Brillouin zone (SBZ) of h-BN is superimposed, and high symmetry points $\bar{\Gamma}, \bar{K}_1$ and $\bar{K}_2$ are marked. Two $K$-points may be distinguished because the mirror symmetry is broken in the presented photoemission experiments and Rh(111) is left with $C_3$ symmetry.
The two maps are very similar. 
This indicates that 1 monolayer of h-BN is transparent for 17 eV photoelectrons, and as for the case of h-BN/Ni(111) no new bands are observed, which confirms the dielectric nature of h-BN nanomesh.
The intensity of the emission is attenuated by the h-BN overlayer.
In Figure \ref{F5}a) the azimuthally averaged intensities from the Fermi surface maps in Figure \ref{F4} are shown as a function of the polar emission angle $\theta$. 
The average intensity of Rh(111), $I_{Rh}$, is always larger than that of h-BN/Rh(111), $I_{hBN}$.
From the attenuation at normal emission $(\theta=0^\circ)$ the scattering cross section $\sigma_{s}$ of a h-BN layer may be estimated for the given electron kinetic energy.
From $\ln(I_{Rh}/I_{hBN})=-\sigma_{s}/A_{uc}$, where $A_{uc}$ is the area of the 1x1 BN unit cell, we get $\sigma_{s}=6.5\pm2 \AA^2$, which is smaller than the theoretical estimate for the elastic scattering cross section \cite{gre02}.

Figure \ref{F5}b) displays the asymmetry $A=(I_{Rh}-I_{hBN})/(I_{Rh}+I_{hBN})$ for the data in Figure \ref{F5}a). 
It shows that not all features of the Fermi surface map are attenuated with the same factor.
Also, it does not scale with a smooth polar angle dependence like e.g. $1/\cos(\theta)$.
The different work functions of 5.50 and 4.14 eV, respectively, impose slightly different cuts across the Fermi surface, and the Fermi surfaces may be distorted by charge transfer (polarization) as it was found for h-BN/Ni(111) \cite {gre02}. 

In Figure \ref{F5}c) azimuthal cuts on a sector of 160$^\circ$ are shown for Rh(111) and h-BN/Rh(111) where two different polar angles $\theta_{Rh}=56^\circ$ and $\theta_{hBN}=52^\circ$ were chosen.
The two angles are selected in such a way that for both azimuthal cuts  $k_\parallel$ corresponds to the $\bar{\Gamma}-\bar{K}$ distance in 1x1 h-BN sheets  (1.7$\AA^{-1}$).
In such momentum distribution curves bands that cross the Fermi level show up as peaks. 
These peaks are attenuated and broadened in the presence of a h-BN overlayer.
The attenuation can be understood by the above discussed scattering, though if this scattering is isotropic, it can not explain the broadening which is in the order of a few degrees.
Here we propose that the broadening is related to the violation of the $k_\parallel$ conservation law in photoemission, if lateral electric fields are present in the vacuum, which is the case for h-BN nanomesh.
For ideal conditions, i.e. no lateral electric fields, we find $k_\parallel = \sin{\theta}\sqrt{2 m_e E_{kin}^V}/\hbar$, where $\theta$ is the measured emission angle, $m_e$ the electron mass, $\hbar$ the Planck constant. $E_{kin}^V$ is the measured electron kinetic energy referred to the vacuum level $E_{kin}^V = \hbar\omega -E_B^F -\Phi$, with $\hbar\omega$ being the photon energy (here = 21.2 eV), $E_B^F$  the binding energy with respect to the Fermi level (here =0), and $\Phi$ the work function.
If the photoelectron current density does not depend on the site within the nanomesh unit cell, the locally different work functions imply a smearing of the emission directions, where this smearing should increase with polar angle $\theta$, and be proportional to the local work function differences.
The increase of the broadening with polar emission angle is indeed seen in the data, and may attain 5 degrees.
If the photoelectron current density depends on the emission site in the unit cell, also shifts of the center of gravity in the emission direction are expected.

Finally, we would like to draw attention on the extra intensity near the $sp$ bands in the second surface Brillouin zone  in the $[\bar{1}\bar{1}2]$ azimuth (see Figure \ref{F4}). They show up as $\bigvee$-shaped intensity near $[\bar{1}\bar{1}2]$.
Clearly, the pattern has $C_{3}$ symmetry, where the mirror symmetry of the (111) surface is broken due to the oblique incidence of the photons.
This rules it out to be caused by the h-BN alone, since the electronic structure of h-BN is expected to be six fold symmetric ($C_{6}$).
It is reminiscent to features as found for the case of h-BN/Ni(111), where it was assigned to the scattering of $sp$ electrons at the interface \cite{gre02}.

In conclusion it is shown that the Fermi surface map of h-BN/Rh(111) gives insight into the electron dynamics at this interface. The comparison with the Fermi surface map of the bare Rh(111) indicates no new Fermi level crossings that could be assigned to a metallic character in h-BN nanomesh and confirms the picture of a corrugated single layer dielectric.
The recently found local work function differences are also reflected in the Fermi surface maps where it is proposed that the laterally varying electric fields  smear out the sharp band crossings from the Fermi surface beneath.  

Fruitful discussions with Sebastian G\"unther are gratefully acknowledged. The project was financially supported by the Swiss National Science Foundation.



\bibliography{references_TG} 


\end{document}